\documentclass[twocolumn,prl,superscriptaddress]{revtex4}
\pdfoutput=1
\usepackage{graphicx}
\usepackage{amsmath}
\usepackage{xcolor}
\usepackage{ifpdf}

\begin{document}
%-------------------------------------------------------------------------
\title{
Experimental Tomographic State Reconstruction of Itinerant Microwave Photons
}
\author{C.~Eichler}
\affiliation{Department of Physics, ETH Z\"urich, CH-8093, Z\"urich, Switzerland.}
\author{D.~Bozyigit}
\affiliation{Department of Physics, ETH Z\"urich, CH-8093, Z\"urich, Switzerland.}
\author{C.~Lang}
\affiliation{Department of Physics, ETH Z\"urich, CH-8093, Z\"urich, Switzerland.}
\author{L.~Steffen}
\affiliation{Department of Physics, ETH Z\"urich, CH-8093, Z\"urich, Switzerland.}
\author{J.~Fink}
\affiliation{Department of Physics, ETH Z\"urich, CH-8093, Z\"urich, Switzerland.}
\author{A.~Wallraff}
\affiliation{Department of Physics, ETH Z\"urich, CH-8093, Z\"urich, Switzerland.}
%------------------------------------------------------------------------
\date{\today}
%-------------------------------------------------------------------------
\begin{abstract}
A wide range of experiments studying microwave photons localized in superconducting cavities have made important contributions to our understanding of the quantum properties of radiation. Propagating microwave photons, however, have so far been studied much less intensely. Here we present measurements in which we reconstruct the Wigner function of itinerant single photon Fock states and their superposition with the vacuum using linear amplifiers and quadrature amplitude detectors. We have developed efficient methods to separate the detected single photon signal from the noise added by the amplifier by analyzing the moments of the measured amplitude distribution up to 4th order. This work is expected to enable studies of propagating microwaves in the context of linear quantum optics.
\end{abstract}
\maketitle

The quantum properties of microwave frequency photons localized in space are typically investigated in the context of cavity quantum electrodynamics (QED) experiments with Rydberg atoms \cite{Haroche2007,Raimond2001} or superconducting circuits \cite{Blais2004,Wallraff2004b,Hofheinz2009}.
Propagating (itinerant) individual microwave photons have been generated \cite{Houck2007} but their quantum properties have not been studied with the same intensity yet.
This is partly due to the fact that efficient single photon counters in this frequency domain are still under development \cite{Chen2010}. However, recently it was shown that characteristic quantum properties of propagating microwave photons, such as antibunching, can be observed in correlation measurements using linear amplifiers and quadrature amplitude detectors \cite{Silva2010,Bozyigit2010a}.
In the context of circuit QED \cite{Blais2004,Wallraff2004b}, propagating microwaves are also used to control \cite{Chow2010} and read out the quantum state of artificial atoms \cite{Palacios-Laloy2010} and to observe phenomena such as resonance fluorescence \cite{Astafiev2010}.

In general, the quantum state of any field mode $a$ is characterized by its density matrix $\rho$ or an equivalent quasi-probability distribution such as the Wigner, the Husimi Q  or the Glauber-Sudarshan P function \cite{Gerry2005,Carmichael1999Book}. Less widely appreciated, the mode $a$ is also equivalently specified by the infinite set of its moments $\langle(a^\dagger)^n a^m\rangle$ \cite{Buzek1996}. In this work we use measurements of such field moments up to 4th order to characterize single photon states.

In the optical domain many experimental techniques exist to reconstruct the quantum state of light using square law detectors \cite{Lvovsky2009}. In many instances, homodyne detection schemes are used to record the statistical properties of single quadrature components for different local oscillator phases which allows to reconstruct the Wigner function by an inverse Radon transformation \cite{Smithey1993}. Alternatively, in heterodyne detection schemes the joint statistics of two conjugate quadrature components described by the Husimi Q distribution are measured \cite{Vogel2009}.
Both techniques thus allow for the full reconstruction of the quantum state of a single field mode.

In the microwave domain linear amplifiers are used to measure the amplitude of the signal instead of its intensity. In this case, one can make use of phase sensitive amplifiers in homodyne detection, ideally amplifying only one quadrature of the signal noiselessly, or of phase insensitive amplifiers in heterodyne detection, amplifying both quadratures of the signal equally while adding at least the vacuum noise to the signal \cite{Caves1982,Clerk2010}.

\begin{figure}[!b]
\centering
\includegraphics[scale=.162]{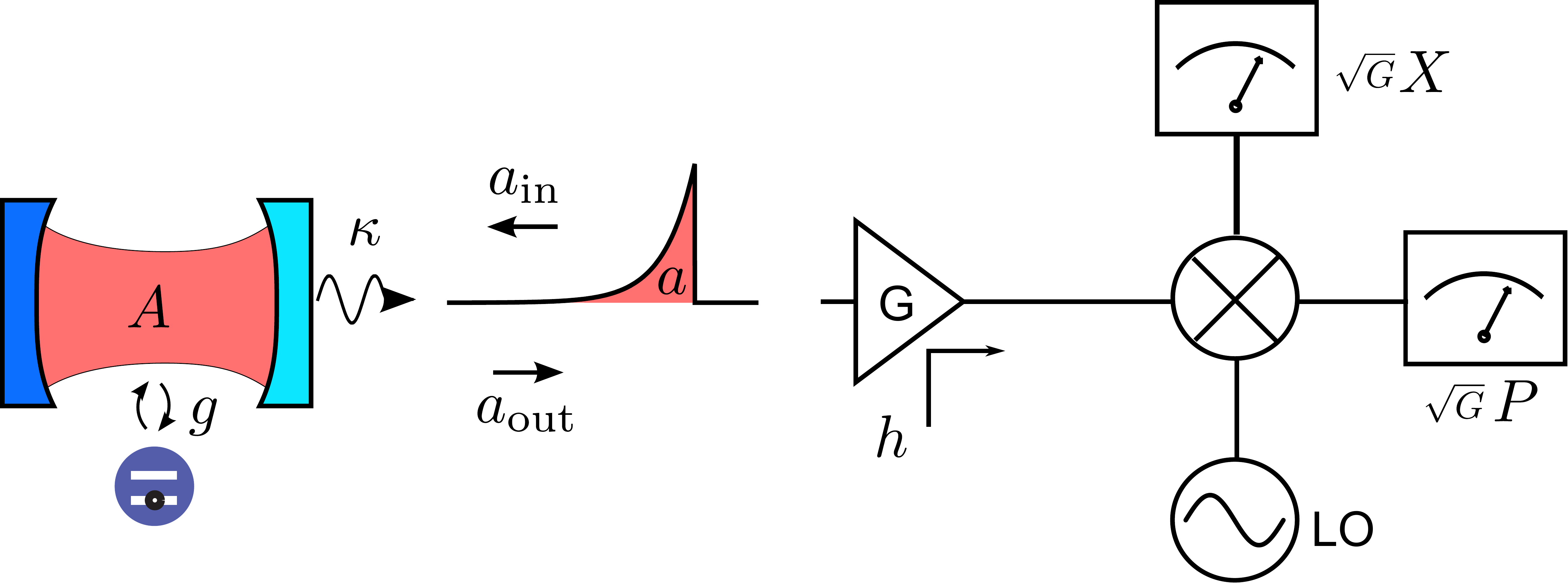}
\caption{Simplified schematic of the experimental setup. An optical analog of the photon source realized in our experiments is shown as a single sided cavity with one highly reflective and one partially transmitting mirror. A single photon is generated in the cavity by preparing the qubit in its excited state when detuned from the cavity and then tuning it into resonance with the cavity for half a vacuum Rabi period $2\pi/g$. The photon is then emitted at a rate $\kappa$ into the output mode $a_{\rm{out}}$ resulting in a exponentially decaying envelope of the single photon pulse while the mode $a_{\rm{in}}$ remains in the vacuum state. The signal is then amplified with effective gain $G$ and noise in mode $h$ is added. The amplified signal is down converted in a microwave quadrature mixer using a local oscillator (LO). The two quadrature amplitudes $X$ and $P$ are recorded using a analog-to-digital converter and stored in real time in a two dimensional histogram using field programmable gate array (FPGA) electronics.
}
\label{fig:setup}
\end{figure}

Here we demonstrate photon state tomography using a phase insensitive amplifier in combination with heterodyne detection of both field quadratures. We first discuss the relation of the single propagating field mode to the resonator mode acting as the photon source (see Fig.~\ref{fig:setup}).
Then we describe a method to systematically separate the relevant signal from the noise using a single linear amplifier chain. Similar methods have recently been discussed for a setup with two linear amplifier channels \cite{Menzel2010}. Finally, we use this method to reconstruct the Wigner function of a single photon Fock state and its superposition with the vacuum state.

We realize a single photon source in a circuit QED setup employing a transmission line resonator of frequency $\nu_r \approx 6.77$~GHz coupled to a single transmon qubit with vacuum Rabi rate $2g/2\pi \approx 146$~MHz \cite{Bozyigit2010a}. By tuning the qubit prepared in a superposition state $\alpha |g\rangle + \beta |e\rangle$ into the resonator for exactly half a vacuum Rabi period we generate single photon states $\alpha |0\rangle + \beta |1\rangle$ \cite{Hofheinz2008, Bozyigit2010a}. The state preparation time is short compared to the cavity decay time $\tau = 1/\kappa \approx 40 \, \rm{ns}$. We repeat the photon generation every $800 \, \rm{ns}$ which allows us to prepare approximately $4 \times 10^9$ single photon states per hour.

The field generated in the resonator then decays into the output mode $a_{\text{out}}$  related to the resonator mode $A$ by the input-output boundary condition $ a_{\text{out}}(t)= \sqrt{\kappa} A(t) - a_{\text{in}}(t)$ \cite{Gardiner1985} where $a_{\text{in}}$ is in the vacuum state, see Fig.~\ref{fig:setup}. We integrate the output signal over a weighted time window $f(t)$ to define a single time independent mode $a = \int \text{d}t f(t) a_{\text{out}}(t)$. Considering the resonator dynamics $ A(t) = e^{-\frac{\kappa t}{2}} A(0) + \sqrt{\kappa} e^{- \frac{\kappa t}{2}}\int_0^{t} \text{d}\tau e^{\frac{\kappa \tau}{2}} a_{\text{in}}(\tau)$  \cite{Walls1994} the choice $f(t)= {\sqrt{\kappa}}e^{- \frac{\kappa t}{2}} \Theta(t)$, where $\Theta(t)$ is the Heaviside step function, leads to the identity $a = A(0)$. Therefore $a$ is in the same quantum state as the resonator at the preparation time $t=0$.

To characterize the state of mode $a$ we first pass the signal through a phase insensitive amplifier chain with effective gain $G$ which introduces an additional noise mode $h$. The amplified signal is then split equally into two parts and mixed with an in-phase and out-of-phase local oscillator, respectively, to simultaneously detect the conjugate quadrature components $\hat{X}$ and $\hat{P}$. Consequently, $\hat{X}$ and $\hat{P}$ are related to $a$ and $h$ by
\begin{equation}
\sqrt{G}(\hat{X} + i \hat{P}) = \sqrt{G}(a + h^\dagger) \equiv \hat{S} \,
\label{eq:SDef}
\end{equation}
where we define the complex amplitude operator $\hat{S}$ \cite{Silva2010}.

If the amplifier is quantum limited, i.e.~the noise mode $h$ is in the vacuum state,
one can show that the quantum state at the amplifier output can be expressed in terms of the Husimi Q function $Q_{\text{out}}(\sqrt{G}\alpha) = \frac{1}{G} Q_{\text{in}}(\alpha)$ \cite{Nha2010}. This implies that the Q function remains invariant under the amplification process up to a scaling factor. Therefore the measurement results of the operator $\hat{S}$  are distributed as $Q_{\text{out}}(S)$.

The best commercially available amplifiers for frequencies below 10 GHz have a noise temperature of about $T_{\text{noise}}=2 \, \rm{K}$ and therefore are far from being quantum limited. In this case, the noise mode $h$ is in good approximation in a thermal state which is represented by a Gaussian phase space distribution.
As shown in Ref.~\cite{Kim1997}, the measured distribution
\begin{equation}
D^{[\rho]}(S) = \frac{1}{G} \int \text{d}^2 \beta P_a(\beta) Q_h (S^*/\sqrt{G}-\beta^*)
\label{eq:ON}
\end{equation}
of $\hat{S}$ at the amplifier output can then be interpreted as the convolution of the $P$ function of mode $a$ and the Q function of the noise mode $h$.

\begin{figure}[!b]
\centering
\includegraphics[scale=1.2]{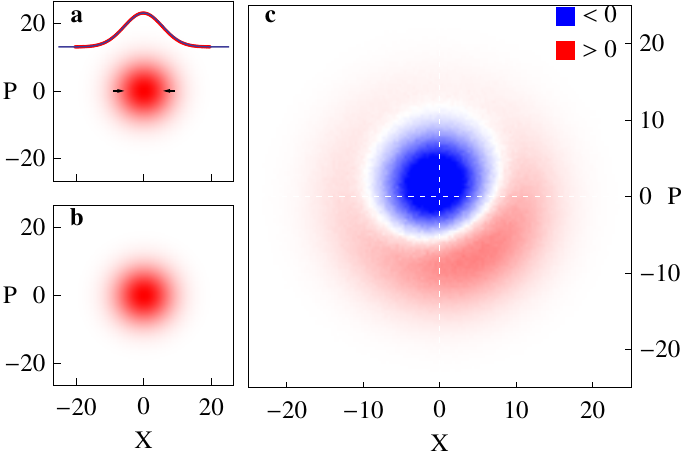}
\caption{({a}) Measured quadrature histogram $D^{[|0 \rangle\langle 0|]}(S)$ for $a$ in the vacuum where $S=\sqrt{G}(X +i P)$. The inset shows a horizontal cut through the histogram (red). The distribution is well described by a normal distribution (blue) with width $\sigma = 5.7$ (indicated by black arrows) corresponding to a system noise temperature of $T_{\rm noise} \approx 21 K$. ({b}) Quadrature histogram $D^{[|1 \rangle\langle 1|]}(S)$ for preparation of single photon Fock states. ({c}) Difference of the two histograms  $D^{[|1 \rangle\langle 1|]}(S)$  and  $D^{[|0 \rangle\langle 0|]}(S)$. }
\label{fig:Data}
\end{figure}

We store the results of repeated measurements of $\hat{S}$ in a two-dimensional histogram with $1024\times1024$ entries which corresponds to a discretized version of the probability distribution $D^{[\rho]}(S)$.
To extract the properties of mode $a$ alone we perform both a measurement in which mode $a$ is left in the vacuum
serving as a reference measurement for the noise where $P_{a}(\beta)=\delta^{(2)}(\beta)$ is a two-dimensional Dirac $\delta$-function
resulting in the distribution \cite{Carmichael1999Book}
\begin{equation}
D^{[|0 \rangle\langle 0|]}(S) = \frac{1}{G} Q_h (S^*/\sqrt{G}),
\label{eq:OFF}
\end{equation}
and a measurement in which the state of interest $|\psi \rangle$, such as a Fock state $|1 \rangle$, is prepared.
The measured histograms for both the vacuum $D^{[|0 \rangle\langle 0|]}$ and for a Fock state $D^{[|1 \rangle\langle 1|]}$ are dominated by the noise added by the amplifier, see Fig.~\ref{fig:Data}({a}) and ({b}).
However, when calculating the numerical difference of both histograms, see Fig.~\ref{fig:Data}({c}), we already clearly observe the circular symmetric character of the single photon
phase space distribution. The small deviation from an ideal circular symmetry is explained  by a slight coherent admixture of the vacuum $|0\rangle$ to the single photon Fock state $|1\rangle$ due to small errors in the state preparation.

\begin{figure}[b]
\centering
\includegraphics[scale=.85]{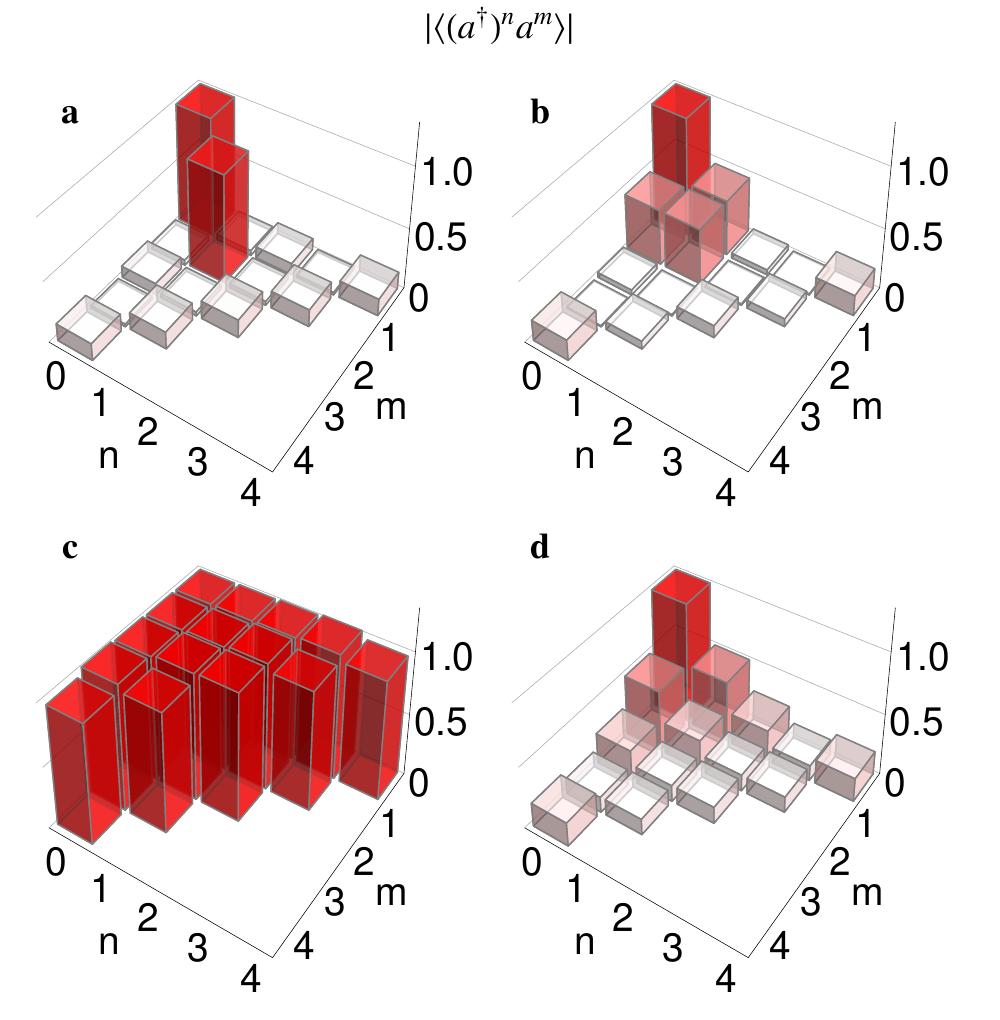}
\caption{Absolute value of the normally ordered moments $|\langle (a^\dagger)^{n} a^{m} \rangle|$  up to 4th order for ({a}) a single photon Fock state,
({b}) a superposition state $(|0\rangle - |1\rangle)/\sqrt{2}$,
and two coherent states with amplitude ({c}) $\alpha = 1$,
and ({d}) $\alpha =0.5$.
}
\label{fig:BarChart}
\end{figure}

To further analyze the data we calculate the moments
\begin{eqnarray}
\langle(\hat{S}^\dagger)^n \hat{S}^m  \rangle_{\rho} &=& \int \text{d}^2 S \,\, (S^*)^n S^m \,\,D^{[\rho]}(S)
\end{eqnarray}
of the two histograms up to a given order that is chosen to be $n+m=4$ in our experiments.
This specific choice will be justified later in the text.
When noise and signal are uncorrelated the calculated moments correspond to the operator averages
\begin{eqnarray}
\langle(\hat{S}^\dagger)^n \hat{S}^m \rangle_{\rho} &=&
\nonumber
\\
& & \hspace{-20mm}
G^{\frac{n+m}{2}}\sum_{i,j=0}^{n,m} \binom{m}{j}\binom{n}{i}
\langle(a^\dagger)^i a^j  \rangle
\langle h^{n-i} (h^\dagger)^{m-j}\rangle,
\label{eq:ON}
\end{eqnarray}
reducing to $\langle(\hat{S}^\dagger)^n \hat{S}^m  \rangle_{|0\rangle \langle 0|} = G^{\frac{n+m}{2}}\langle h^{n} (h^\dagger)^{m}\rangle \,\,$
when $a$ is in the vacuum state. Equation~(\ref{eq:ON}) can then be inverted to calculate the moments $\langle (a^\dagger)^{n} a^{m} \rangle$ from $ \langle(\hat{S}^\dagger)^n \hat{S}^m \rangle_{\rho}$ and $\langle(\hat{S}^\dagger)^n \hat{S}^m  \rangle_{|0\rangle \langle 0|}$ up to the desired order, 4 in this case,
as shown in Fig.~\ref{fig:BarChart}({a}).
\begin{figure}[b]
\centering
\includegraphics[scale=.5]{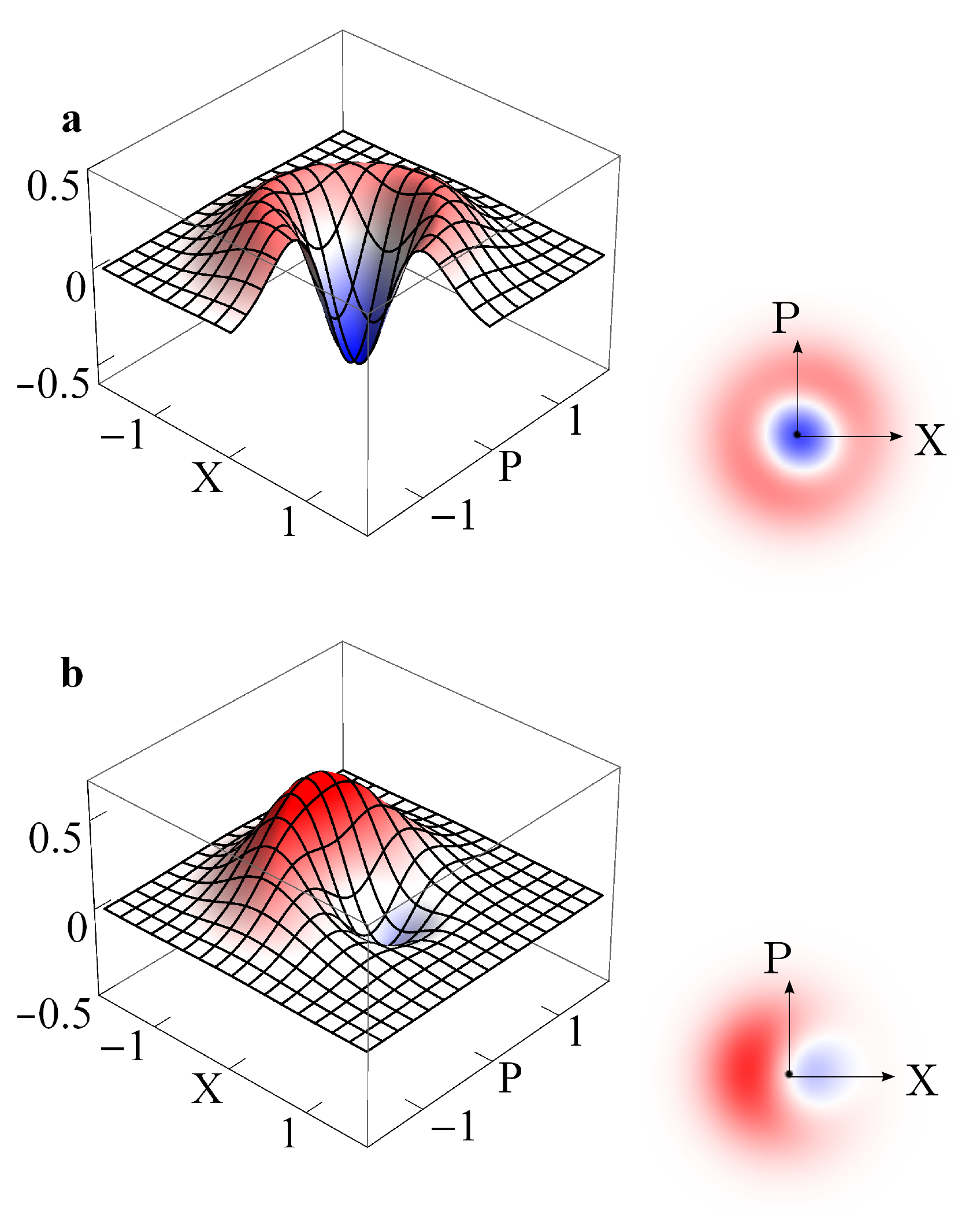}
\caption{Wigner function $W(\alpha = X + i P)$ for ({a}) a single photon Fock state and ({b}) a superposition state both reconstructed from the measured moments shown in Fig.~\ref{fig:BarChart}.}
\label{fig:Wigner}
\end{figure}
We note that the quadrature histograms are normalized such that the zeroth order moments are always unity for all prepared states. The off diagonal elements in the moment matrix express coherences between different photon number states. They vanish for states with circular symmetric phase space distributions such as pure Fock states or thermal states. For the Fock state $|1\rangle$ (Fig.~\ref{fig:BarChart}(a)), we observe that all off diagonal moments are close to zero. In addition, we note that the fourth order moment $\langle(a^\dagger)^2 a^2\rangle$ is also close to $0$ indicating antibunching of the prepared single photon states \cite{Bozyigit2010a}. In contrast, a thermal state with the same mean photon number would also have vanishing off diagonal moments but finite diagonal 4th order moments.
Experimentally, for the single photon Fock state, the aforementioned residual coherent admixture of the vacuum state leads to a non-vanishing small mean amplitude $|\langle a \rangle| = 0.044$ and a slightly reduced mean photon number $\langle a^\dagger a \rangle = 0.91$. For an integration time of 12 hours for each state, we find errors of the 4th order moments to be approximately $\pm 0.1$ where the statistical error in the moments is known to increase exponentially with increasing order \cite{Silva2010}.
In comparison, the estimated statistical errors for the first, second and third order moments are approximately $1.5 \times 10^{-3}$, $4.5 \times 10^{-3}$, and $1.5 \times 10^{-2}$, respectively.

We have also prepared and analyzed superposition states of the type $(|0\rangle + e^{i \phi} |1\rangle)/\sqrt{2}$, see Fig.~\ref{fig:BarChart}(b). The relative phase $\phi$ is controlled by the phase of the corresponding qubit state that is mapped into the resonator. For this class of states, the mean amplitude ideally equals the mean photon number $|\langle a \rangle| = \langle a^\dagger a \rangle = 0.5$.
The first equality remains valid even if the state is slightly mixed with the vacuum.
We have been able to use this property to determine the effective gain $G$ of our amplifier chain because first and second order moments have a different characteristic scaling with $G$. This allowed us to scale $X$ and $P$ axes of the histograms (Fig.~\ref{fig:Data}) such that they correspond exactly to real and imaginary part of $a + h^\dagger$. From our measurement data, we extract $|\langle a \rangle|=0.466$ which is close to the expected value.

To further confirm the validity of our scheme, we have generated coherent states $|\alpha\rangle$ with amplitude $\alpha = 1$ and  $\alpha = 0.5$ by applying $10 \, \rm{ns}$ square coherent pulses with controlled amplitude to the weakly coupled input port of the resonator. The moments of coherent states are given by $\langle (a^\dagger)^n a^m\rangle=(\alpha^*)^n\alpha^m$. For $\alpha=1$ all moments are observed to be close to $1$ (Fig.~\ref{fig:BarChart}({c})), as expected.
This also demonstrates that systematic errors in the detection chain, such as small nonlinearities, are negligible as all moments take their expected values.
For $\alpha=0.5$ (Fig.~\ref{fig:BarChart}({d})), the measured moments decay exponentially with $\langle (a^\dagger)^n a^m\rangle=0.5^{n+m}$, as expected.
The fourth order moments appear larger than the third order ones, due to their larger statistical error.

From the measured moments we have reconstructed the Wigner function $W(\alpha)$ for a single photon Fock state and its superposition with the vacuum (Fig.~\ref{fig:Wigner}).
It is sufficient to evaluate \cite{Haroche2007}
\begin{eqnarray}
W(\alpha)= \sum_{n,m}
\int \text{d}^2\lambda \,   \frac{\langle(a^\dagger)^n a^m\rangle (\lambda^*)^n \lambda^m}{ \pi^2 n!m!}    \,e^{-\frac{1}{2}|\lambda|^2+\alpha\lambda^{*}-\alpha^{*}\lambda}
\nonumber
\\
\label{eq:Wigner}
\end{eqnarray}
up to order $n+m = 3$ because $\langle(a^\dagger)^2 a^2\rangle \sim 0$. In general, all higher order moments with $n+ m \geq 2 N -1$ have to be zero if one diagonal moment $\langle (a^\dagger)^N a^N \rangle$ vanishes, which follows from the fact that diagonal moments $\langle k|(a^\dagger)^n a^n|k\rangle$ with $n>k$ are zero for Fock states $|k\rangle$.

The Wigner function of the single photon Fock state Fig.~\ref{fig:Wigner}(a) shows clear negative values which indicate the quantum character of the observed state. The slight shift of $|\langle a \rangle| = 0.044$ from the phase space origin that we already observed in the raw measurement data (Fig.~\ref{fig:Data}c) of the $|1\rangle$ state is also apparent in the reconstructed Wigner function. The superposition state $(|0\rangle -|1\rangle)/\sqrt{2}$  displayed in Fig.\ref{fig:Wigner}(b) has a finite mean amplitude which leads to the finite center of mass of the distribution. Still, negative values in the distribution persist, illustrating the quantum coherence between the $|0\rangle$ and $|1\rangle$ state.
We have also varied the relative phase $\phi$ of the superposition states and have observed the according rotation of the Wigner function (data not shown).

In summary, we have reconstructed the Wigner function of itinerant single microwave photons and small coherent fields using linear amplification, quadrature amplitude detection and efficient data analysis even in the presence of noise added by the amplifier. We have implemented a method to separate the quantum signal from the amplifier noise in a measurement setup with only one detection channel. We believe that propagating microwaves will be investigated more intensely in the context of future quantum optics and also quantum information processing experiments \cite{Kok2007} where low noise parametric amplifiers \cite{Castellanos2007,Bergeal2010,Yamamoto2008} have the potential to significantly improve the detection efficiency.

\bibliographystyle{apsrev}
%-----------------------------------------------------
\begin{acknowledgments}
The authors would like to acknowledge fruitful discussions with Alexandre Blais, Marcus da Silva, Gerard Milburn and Barry Sanders. This work was supported by
the European Research Council (ERC) through a Starting Grant and by ETHZ.
\end{acknowledgments}
%-------------------------------------------------------------------------
\bibliographystyle{apsrev}

\end{document}